\documentclass[
  english,
  techrep
]{ipsj}

\usepackage{my_packages}
\usepackage{my_command_definitions}

\usepackage[varg]{txfonts}
\makeatletter%
\input{ot1txtt.fd}
\makeatother%

\begin{document}

\title{Practical Persistent Multi-Word Compare-and-Swap Algorithms for Many-Core CPUs}

\affiliate{NU}{%
  Graduate School of Informatics, Nagoya University,
  Nagoya-shi, Aichi 464-8601 Japan%
}

\author{Kento Sugiura}{NU}[sugiura@i.nagoya-u.ac.jp]
\author{Manabu Nishimura}{NU}
\author{Yoshiharu Ishikawa}{NU}

\begin{abstract}
  In the last decade, academic and industrial researchers have focused on \emph{persistent memory} because of the development of the first practical product, Intel Optane.
  One of the main challenges of persistent memory programming is to guarantee consistent durability over separate memory addresses, and Wang et al. proposed a \emph{persistent multi-word compare-and-swap (PMwCAS) algorithm} to solve this problem.
  However, their algorithm contains redundant compare-and-swap (CAS) and cache flush instructions and does not achieve sufficient performance on many-core CPUs.
  This paper proposes a new algorithm to improve performance on many-core CPUs by removing useless CAS/flush instructions from PMwCAS operations.
  We also exclude dirty flags, which help ensure consistent durability in the original algorithm, from our algorithm using PMwCAS descriptors as write-ahead logs.
  Experimental results show that the proposed method is up to ten times faster than the original algorithm and suggests several productive uses of PMwCAS operations.
\end{abstract}


\maketitle
\thispagestyle{plain}
\pagestyle{plain}

\section{Introduction}

\emph{Persistent (non-volatile) memory} is one of the most impressive hardware developments of the last decade.
Persistent memory is byte-addressable and comparable to volatile memory while guaranteeing sufficient storage space and the durability of written data.
These characteristics allow software developers to control their data with a finer and more flexible read/write unit, which is 256 bytes in the case of Intel Optane.
As a result, academic and industrial researchers have proposed various middleware and libraries, such as databases~\cite{PVLDB:Gong2022,PVLDB:Liu2021,PVLDB:Yan2021}, hash tables~\cite{PVLDB:Vogel2022,SIGMOD:Hu2022,PVLDB:Hu2021}, and tree-structured indexes~\cite{PVLDB:Zhang2022a,PVLDB:Zhang2022b,PVLDB:He2022,PVLDB:Lersch2019}.
Although Intel Optane, the first persistent memory product, is no longer in production, vendors have continued to develop technologies such as Storage Class Memory (SCM) and Compute Express Link (CXL) that will reach persistent memory~\cite{SCW:Fridman2023}.

\begin{figure}[tb]
  \centering
  \includegraphics{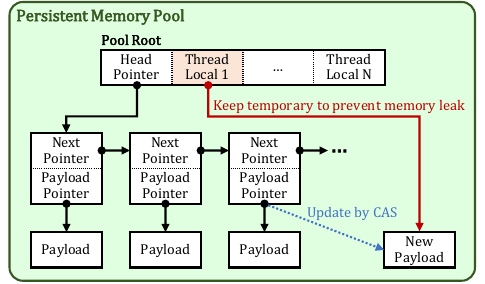}
  \caption{Updating a payload pointer in a persistent linked list.}
  \label{fig:linked-list}
\end{figure}

One of the main challenges of using persistent memory is to ensure consistent durability.
When developers construct data structures with traditional storage, temporary workspaces (i.e., volatile memory) and permanent storage (i.e., SSD or HDD) are separated.
This separation allows developers to control data durability through system calls.
On the other hand, developers must be careful about the persistence states of their data because \emph{persistent memory acts as both workspaces and permanent storage}.
For example, we consider updating a data pointer in persistent memory.
\figref{fig:linked-list} shows that each node of a linked list has a pointer to its payload, and a worker thread attempts to update it to a new pointer.
Note that we assume that the Persistent Memory Development Kit (PMDK)~\cite{URL:PMDK} is used to manage persistent memory;  persistent memory is divided into a pool that has a root region as a starting point for data access.
When a thread updates a payload pointer, the change is initially reflected in CPU caches and \emph{is not persistent}.
Therefore, developers must control data durability through \emph{cache flush instructions} and prepare appropriate recovery procedures.
In this example, since incomplete payload updates may exist after a machine failure, a recovery procedure must read thread-local regions and release allocated payloads.
Although both traditional storage and persistent memory require explicit data persistence, a cache controller flushes caches at unexpected times and complicates programming in persistent memory.

To support programming in persistent memory, Wang et~al. proposed \emph{persistent multi-word compare-and-swap (PMwCAS) operations}~\cite{ICDE:Wang2018}.
The above example still has a problem; the recovery procedure may inadvertently release embedded payloads and fail to release old payloads.
After worker threads have updated and flushed their payloads, their thread-local regions still contain them, and there is no reference to old payloads in a linked list.
If a machine crashes at such a time, the recovery procedure cannot restore a consistent state.
PMwCAS operations can solve this problem by \emph{swapping multiple words simultaneously}; a thread can atomically update a payload pointer in a node and a temporary pointer in a thread-local region (i.e., swap old/new payload pointers).
Consequently, a thread-local region holds an old payload after updates, and the recovery procedure can release it when a machine crashes.

\begin{figure}[tb]
  \centering
  \includegraphics{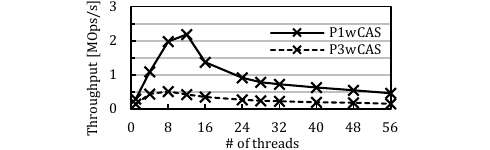}
  \caption{Throughput of Wang et al.'s persistent one/three-word CAS operations in high-competitive environments.}
  \label{fig:original-pmwcas-over-thread}
\end{figure}

In this paper, we propose new PMwCAS algorithms for many-core CPUs.
Although PMwCAS operations help to program in persistent memory, the original algorithm has a performance problem, as shown in \figref{fig:original-pmwcas-over-thread}.\footnote{%
  The detailed experimental settings are shown in \Sec{\ref{sec:experiments}}.
}
The figure shows the throughput of PMwCAS operations per second and its inefficiency on many cores.
This performance degradation results from frequent cache invalidations due to redundant CAS and flush instructions.
Since we have already identified this problem and proposed an improved MwCAS algorithm~\cite{IEICE:Sugiura2022}, this paper extends it to persistent memory.
We summarize our contributions as follows.
\begin{itemize}
  \item We remove redundant CAS and flush instructions in our PMwCAS algorithms.
        In addition to the improvements applied to our MwCAS operations~\cite{IEICE:Sugiura2022}, we exclude \emph{dirty flags} that manage data durability in the original algorithm.
  \item We implement the proposed methods as a C++ library~\cite{URL:pmem-atomic}.
        The implementation has no external dependencies except for the PMDK.
        Unlike the original algorithm, our algorithms do not require garbage collection and achieve stable behavior as a library.
  \item Experimental results demonstrate the effectiveness of the proposed methods.
        Our PMwCAS algorithms are up to ten times faster than the original algorithm.
        We also evaluate performance trends according to several parameters and summarize suggestions for handling of PMwCAS operations.
\end{itemize}

The rest of this paper is organized as follows.
In \Sec{\ref{sec:related-work}}, we present related work.
We introduce a new PMwCAS algorithm with dirty flags in \Sec{\ref{sec:pmwcas-with-df}}, and then we exclude dirty flags from PMwCAS operations in \Sec{\ref{sec:pmwcas-without-df}}.
We experimentally evaluate the proposed methods in \Sec{\ref{sec:experiments}}, and we conclude the paper in \Sec{\ref{sec:conclusion}}.

\section{Related Work}
\label{sec:related-work}

This section explains existing multi-word compare-and-swap (MwCAS) algorithms and provides an overview of programming in persistent memory environments.
In the following of this paper, we assume that a CAS instruction works as in \figref{fig:algo:cas}.

\begin{algorithm}[tb]
  \DontPrintSemicolon
  \SetKwRepeat{Do}{do}{while}
  \SetKw{Break}{break}
  \KwIn{$address$\tcp*[r]{a target memory address}}
  \KwIn{$expected$\tcp*[r]{an expected (old) value}}
  \KwIn{$desired$\tcp*[r]{a desired (new) value}}
  \KwOut{$word$\tcp*[r]{a current value}}

  $word \leftarrow$ load a word from $address$\;
  \If{$word = expected$}{
    store $desired$ into $address$
  }

  \caption{An algorithm assumed herein to be a CAS instruction.}
  \label{fig:algo:cas}
\end{algorithm}

\subsection{Multi-Word Compare-and-Swap Algorithms}

MwCAS operations are procedures for atomically swapping multiple words\footnote{%
  We assume that each word is represented by eight bytes in this paper.
}~\cite{DISC:Harris2002,IJPP:Sundell2011,IJPP:Feldman2015,DISC:Guerraoui2020,IEICE:Sugiura2022}.
Harris et~al. proposed the first practical MwCAS algorithm, the CASN algorithm, which can be used in commodity machines without any specific hardware support~\cite{DISC:Harris2002}.
Sundell~\cite{IJPP:Sundell2011} and Feldman et~al.~\cite{IJPP:Feldman2015} proposed wait-free MwCAS algorithms to prevent worker thread starvation.
However, as with other wait-free algorithms~\cite{BOOK:Herlihy2021ch3}, the experimental results showed that wait-free features do not provide significant performance improvements.
Guerraoui et~al. proposed the AOPT algorithm to reduce the number of CAS instructions needed for each MwCAS~\cite{DISC:Guerraoui2020}.
However, this approach requires additional garbage collection and may reduce performance in some workloads.

Our previous work excluded any garbage collection from MwCAS operations and reduced the number of CAS instructions~\cite{IEICE:Sugiura2022}.
While the CASN and AOPT algorithms~\cite{DISC:Harris2002,DISC:Guerraoui2020} require $3k$ and $2k + 1$ CAS instructions, respectively, for a $k$-word CAS operation, our algorithm can perform it with $2k$ CAS instructions in ideal cases (i.e., when there are no conflicts).
Besides, the existing PMwCAS algorithm follows the CASN algorithm and requires $4k$ CAS instructions~\cite{ICDE:Wang2018}.
The experimental results showed that our algorithm outperformed the existing PMwCAS implementation in volatile memory~\cite{IEICE:Sugiura2022} and motivated us to extend this algorithm to persistent memory.

We briefly explain the core concepts of MwCAS operations~\cite{DISC:Harris2002,IEICE:Sugiura2022}.
MwCAS operations reserve target addresses by embedding MwCAS descriptors there.
If an MwCAS operation has reserved all target addresses, the operation succeeds and swaps to new values.
If an MwCAS operation fails to reserve even one address, the operation fails and reverts to old values.
Developers must be careful about the order of embedding because this embedding process also acts as a linearization of MwCAS operations.
In other words, if MwCAS operations embed their descriptors in any order, it will cause corrupted MwCAS states such as deadlocks.

\subsection{Programming with Persistent Memory}

Developing software that functions correctly is a primary challenge in using persistent memory.
Developers must be careful when writing programs to avoid inconsistent software states because persistent memory ensures that all flushed data are durable.
For example, if developers mishandle pointers to allocated memory regions, it leads to persistent memory leaks~\cite{BOOK:Scargall2020ch4}.
Thus, academic and industrial researchers have proposed middleware and libraries, such as OLTP engines~\cite{PVLDB:Yan2021,SIGMOD:Arulraj2017}, tree-structured indexes~\cite{PVLDB:Lersch2019,PVLDB:He2022}, and hash tables~\cite{PVLDB:Hu2021}, to provide programmers with high-level APIs.
However, developers of these middleware and libraries still need to carefully program for persistent memory.

The Persistent Memory Development Kit (PMDK)~\cite{URL:PMDK} is a collection of libraries and tools that support programming in persistent memory.
A transaction API in the PMDK provides programmers with atomicity, consistency, isolation, and durability (i.e., the ACID property) for their software~\cite{BOOK:Scargall2020ch4}.
When programmers wrap their critical sections in transactions, the PMDK guarantees ACID when processing them.
However, since the PMDK achieves transaction processing using lock-based concurrency controls, the transaction API can degrade the performance of low-level middleware and libraries.

Persistent CAS (PCAS) operations have been proposed in both software~\cite{ICDE:Wang2018} and hardware~\cite{SPAA:Wang2019} domains to support the implementation of lock-free data structures in persistent memory.
Since CAS instructions are one of the essential components of lock-free algorithms~\cite{BOOK:Herlihy2021ch5}, PCAS operations are also crucial for achieving lock-free and durable data structures.
Wang et~al.~\cite{ICDE:Wang2018} extended PCAS to PMwCAS operations to support the implementation of more complicated data structures, such as a lock-free \BTree{}~\cite{PVLDB:Arulraj2018}, by simultaneously swapping multiple words.

\section{PMwCAS Algorithm with Dirty Flags}
\label{sec:pmwcas-with-df}

First, we introduce a new PMwCAS algorithm with \emph{dirty flags}.
A dirty flag indicates that a target word is visible in CPU caches but has not yet been persisted (i.e., flushed to persistent memory).
If worker threads read such unflushed words and continue their processes, it can lead to corrupted results after machine failures due to inconsistency between caches and persistent memory.
Thus, worker threads must 1) flush dirty words before continuing their processes or 2) wait for words to be flushed.
Although Wang et al. have adopted the first approach~\cite{ICDE:Wang2018}, this approach leads to frequent small writes and double flushes, which have been reported as potential bottlenecks in the existing work~\cite{PVLDB:Gugnani2020,PVLDB:Benson2022}.
In addition, our previous work has shown that such an approach causes frequent cache invalidations and degrades the performance of MwCAS~\cite{IEICE:Sugiura2022}.
Based on these experimental results, we adopt the second approach in this paper.

According to the existing work~\cite{DISC:Harris2002,ICDE:Wang2018,IEICE:Sugiura2022}, we use a \emph{descriptor} in \tabref{tab:descriptor} to manage each PMwCAS operation.
A PMwCAS descriptor has the state of a corresponding PMwCAS operation (Failed, Succeeded, or Completed) and the information to perform it.
Target information contains a destination address, an expected value, and a desired value to perform each CAS instruction.\footnote{%
  Although our implementation can hold a memory barrier in each target information, we omit its details here because its usage depends on a constructed data structure.
}
Since a descriptor also acts as a write-ahead log, we can roll forward/back a corresponding PMwCAS operation after machine failures.

\begin{table}[tb]
  \caption{A PMwCAS descriptor.}
  \label{tab:descriptor}
  \centering \small
  \begin{tabular}{ll}
    \toprule
    Field name     & Content                                  \\
    \cmidrule(r){1-1}\cmidrule(l){2-2}
    state          & The current state of a PMwCAS operation. \\
    count          & The number of CAS operations.            \\
    targets[count] & An array of target CAS information.      \\
    - address      & A destination logical address.           \\
    - expected     & An expected value before CAS.            \\
    - desired      & A desired value after CAS.               \\
    \bottomrule
  \end{tabular}
\end{table}

We show a proposed PMwCAS algorithm and a read procedure in \figref{fig:algo:pmwcas} and \figref{fig:algo:read-pmwcas-field}, respectively.
Note that we use a period to denote accessing a field of a variable (e.g., $var.field$).
The algorithm has reservation and finalization (commit or abort) phases.
First, a worker thread embeds the address of a descriptor to reserve each target address of PMwCAS (lines 4-14).
If a thread cannot reserve all target addresses, the operation has failed and shifts to the abort phase (lines 8-10).
If a thread has reserved and persisted all target addresses, the operation has succeeded and moves to the commit phase (lines 11-14).
In the finalization phase (lines 15-23), a thread uses dirty flags (lines 18-20) and avoids inconsistent states, as described above.
Our read procedure avoids reading intermediate states and waits for any PMwCAS operation to finish.

\begin{algorithm}[tb]
  \DontPrintSemicolon
  \SetKwRepeat{Do}{do}{while}
  \SetKw{Break}{break}
  \KwIn{$desc$\tcp*[r]{a PMwCAS descriptor}}
  \KwOut{$success$\tcp*[r]{true if PMwCAS succeeds}}

  $desc.state \leftarrow \mathtt{Failed}$\;
  persist $desc$\;
  $success \leftarrow \mathtt{true}$\;
  \ForEach(\tcp*[f]{reserve every target address}){$t \in desc.targets$}{
    \Do(\tcp*[f]{wait if another PMwCAS is in progress}){word {\upshape is a PMwCAS descriptor or includes a dirty flag}}{
      $word \leftarrow \CAS(t.address, t.expected,$ address of $desc)$
    }
    \If{$word \neq t.expected$}{
      $success \leftarrow \mathtt{false}$\;
      \Break
    }
  }
  \If{$success$}{
    \ForEach{$t \in desc.targets$}{
      persist the embedded address in $t.address$
    }
    $desc.state \leftarrow \mathtt{Succeeded}$\;
    persist $desc$\tcp*[r]{a linearization point}
  }
  \ForEach(\tcp*[f]{commit or abort PMwCAS}){$t \in desc.targets$}{
    \If{t.address {\upshape does not have a descriptor}}{\Break}
    $word \leftarrow$ \lIf{$success$}{$t.desired$ \textbf{else} $t.expected$}
    \If(\tcp*[f]{skipped in Section 4}){\upshape dirty flags are used}{
      store $word$ with a dirty flag into $t.address$\;
      persist dirty $word$ in $t.address$
    }
    store $word$ into $t.address$\;
    persist $word$ in $t.address$
  }
  $desc.state \leftarrow \mathtt{Completed}$\;

  \caption{A PMwCAS algorithm with/without dirty flags.}
  \label{fig:algo:pmwcas}
\end{algorithm}

\begin{algorithm}[tb]
  \DontPrintSemicolon
  \SetKwRepeat{Do}{do}{while}
  \SetKw{Break}{break}
  \KwIn{$address$\tcp*[r]{a target address}}
  \KwOut{$word$\tcp*[r]{a current value}}

  \Do(\tcp*[f]{wait if any PMwCAS is in progress}){word {\upshape is a PMwCAS descriptor or includes a dirty flag}}{
    $word \leftarrow$ load a word from $address$
  }

  \caption{A read procedure for PMwCAS target addresses.}
  \label{fig:algo:read-pmwcas-field}
\end{algorithm}

\textbf{Implementation Details:}
Our implementation uses the last two bits in each word to distinguish payloads, descriptors, and dirty flag, as shown in \tabref{tab:bit-flags}.
In other words, target words cannot use the last two bits to represent their payloads.
Although this is a restriction of our PMwCAS algorithm, the existing algorithm requires three bits~\cite{ICDE:Wang2018} and CAS targets usually allow this restriction (e.g., tuple IDs and logical addresses).
In addition, we use the test-and-test-and-set (TTAS) algorithm~\cite{BOOK:Herlihy2021ch7} when embedding descriptors to avoid useless cache invalidations.
We also use back-off~\cite{BOOK:Herlihy2021ch7} when reading target addresses to prevent busy loops.

\begin{table}[tb]
  \caption{The last two bits of each word and its content.}
  \label{tab:bit-flags}
  \centering \small
  \begin{tabular}{ll}
    \toprule
    Bit state   & Content                      \\
    \cmidrule(r){1-1}\cmidrule(l){2-2}
    \texttt{00} & A target payload.            \\
    \texttt{10} & A PMwCAS descriptor.         \\
    \texttt{01} & A payload with a dirty flag. \\
    \bottomrule
  \end{tabular}
\end{table}

\begin{figure}[tb]
  \centering
  \includegraphics{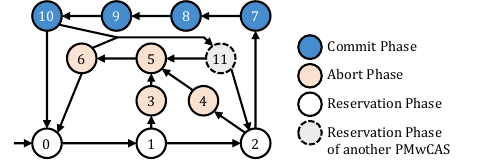}
  \caption{A state machine of PMwCAS operations with dirty flags.}
  \label{fig:state-machine-with-dirty-flag}
\end{figure}

\textbf{Consistency:}
We show a state machine of our PMwCAS algorithm in \figref{fig:state-machine-with-dirty-flag} and a corresponding ID list in \tabref{tab:states-with-dirty-flag} to demonstrate the consistency achieved by the proposed method.
Once a worker thread embeds a descriptor into CPU caches (i.e., after entering state 1), other threads cannot modify the same address in our algorithm (i.e., only the thread traverses among states 1-10).
Thus, we first describe the consistency in states 1-10.
We do not need a recovery process for states 1 and 3 because they have clean values in persistent memory.
Although states 2, 4, 7, and 8 contain a descriptor in persistent memory and require recovery after machine failures, we can achieve it by using target information in an embedded descriptor.
That is, we can roll back/forward to the expected/desired values based on the PMwCAS state of a descriptor (Failed or Succeeded).
In the same way, we can recover states 5, 6, 9, and 10 from intermediate values by clearing dirty flags.
When a PMwCAS operation reaches the end of the abort/commit phase (i.e., state 6 or 10), another thread can swap a word to its descriptor $desc^{*}$ on CPU caches (i.e., moving to state 11).
State 11 has dirty values in persistent memory and is in the reservation phase of another PMwCAS, but we can consistently recover them by clearing dirty flags.
Since a state machine switches to that of another PMwCAS (i.e., moving to state 2 or 5), we can guarantee consistency in the same way.

\begin{table}[tb]
  \caption{IDs of a state machine in \figref{fig:state-machine-with-dirty-flag}. Prime and asterisk symbols denote dirty values and another PMwCAS descriptor, respectively. State column contains the initial of Completed (C), Failed (F), and Succeeded (S).}
  \label{tab:states-with-dirty-flag}
  \tabcolsep=1.5truemm

  \begin{minipage}{0.49\columnwidth}
    \centering \small
    \begin{tabular}{rccc}
      \toprule
      ID & Cache      & PMEM       & State \\
      \cmidrule(r){1-1}\cmidrule(l){2-4}
      0  & $v_{old}$  & $v_{old}$  & C     \\
      1  & $desc$     & $v_{old}$  & F     \\
      2  & $desc$     & $desc$     & F     \\
      3  & $v'_{old}$ & $v_{old}$  & F     \\
      4  & $v'_{old}$ & $desc$     & F     \\
      5  & $v'_{old}$ & $v'_{old}$ & F     \\
      \bottomrule
    \end{tabular}
  \end{minipage}
  \hfill
  \begin{minipage}{0.49\columnwidth}
    \centering \small
    \begin{tabular}{rccc}
      \toprule
      ID & Cache      & PMEM           & State \\
      \cmidrule(r){1-1}\cmidrule(l){2-4}
      6  & $v_{old}$  & $v'_{old}$     & F     \\
      7  & $desc$     & $desc$         & S     \\
      8  & $v'_{new}$ & $desc$         & S     \\
      9  & $v'_{new}$ & $v'_{new}$     & S     \\
      10 & $v_{new}$  & $v'_{new}$     & S     \\
      11 & $desc^{*}$ & $v'_{old/new}$ & F/S   \\
      \bottomrule
    \end{tabular}
  \end{minipage}
\end{table}

\section{PMwCAS Algorithm without Dirty Flags}
\label{sec:pmwcas-without-df}

Dirty flags can help ensure consistency but reduce PMwCAS performance.
Since worker threads must store the same values twice to set and clear dirty flags, it leads to frequent cache invalidations.
Besides, the existing work~\cite{PVLDB:Benson2022} has reported that double flushes degrade performance unless the cache line writeback (CLWB) instruction~\cite{BOOK:Scargall2020ch2} is used.
Since commodity CPUs do not fully support this instruction, the effect of double flushes is not yet negligible.\footnote{%
  In the case of Intel Xeon, the third or more generations fully support the CLWB instruction.
}

To overcome this problem, we propose a PMwCAS algorithm without dirty flags.
The algorithm  is similar to the previous one but omits lines 18-20 in \figref{fig:algo:pmwcas} to avoid setting/clearing dirty flags.
As we explain the consistency of this algorithm in the next paragraph, we introduce a brief idea for removing dirty flags.
The main point is that \emph{PMwCAS descriptors also act as write-ahead logs in recovery}.
Although Wang et al. use dirty flags to ensure data durability, embedded descriptors can play a role because they have sufficient information for recovery.
As shown in \figref{fig:algo:pmwcas}, since descriptors must be embedded and persisted before dirty flags are set, we can safely exclude dirty flags from our algorithm.

\begin{figure}[tb]
  \centering
  \includegraphics{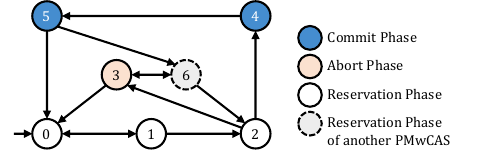}
  \caption{A state machine of PMwCAS operations without dirty flags.}
  \label{fig:state-machine-without-dirty-flag}
\end{figure}

\textbf{Consistency:}
We use a state machine again to demonstrate the consistency achieved by our algorithm without dirty flags.
\figref{fig:state-machine-without-dirty-flag} and \tabref{tab:states-without-dirty-flag} show a state machine of the algorithm without dirty flags and a corresponding ID list.
We can guarantee the consistency of states 1-5 in the same way as the previous section.
State~1 has clean values in persistent memory and does not need recovery.
States 2-5 contain descriptor addresses in persistent memory, and we can recover them by using target information in the corresponding descriptors.
Although state 6 enters the reservation phase of another PMwCAS (i.e., $desc^{*}$) and has previous descriptor addresses (i.e., $desc$) in persistent memory, we can still recover it by the previous descriptors.
In other words, since the expected values of another PMwCAS (i.e., $v^{*}_{old}$) are the same as the reverted/swapped ones of the previous PMwCAS (i.e., $v_{old}$ or $v_{new}$), the recovery results are consistent.

\begin{table}[tb]
  \caption{IDs of a state machine in \figref{fig:state-machine-without-dirty-flag}. We use the same symbols in \tabref{tab:states-with-dirty-flag}}
  \label{tab:states-without-dirty-flag}
  \tabcolsep=1.5truemm

  \begin{minipage}{0.49\columnwidth}
    \centering \small
    \begin{tabular}{rccc}
      \toprule
      ID & Cache     & PMEM      & State \\
      \cmidrule(r){1-1}\cmidrule(l){2-4}
      0  & $v_{old}$ & $v_{old}$ & C     \\
      1  & $desc$    & $v_{old}$ & F     \\
      2  & $desc$    & $desc$    & F     \\
      3  & $v_{old}$ & $desc$    & F     \\
      \bottomrule
    \end{tabular}
  \end{minipage}
  \hfill
  \begin{minipage}{0.49\columnwidth}
    \centering \small
    \begin{tabular}{rccc}
      \toprule
      ID & Cache      & PMEM   & State \\
      \cmidrule(r){1-1}\cmidrule(l){2-4}
      4  & $desc$     & $desc$ & S     \\
      5  & $v_{new}$  & $desc$ & S     \\
      6  & $desc^{*}$ & $desc$ & F/S   \\
      \bottomrule
    \end{tabular}
  \end{minipage}
\end{table}

\section{Experiments}
\label{sec:experiments}

We evaluate the proposed PMwCAS algorithms by using synthetic datasets.
We implemented the proposed methods and a benchmarking program in C++~\cite{URL:pmem-atomic,URL:pmwcas-bench}.
\tabref{tab:environment} summarizes the experimental environment.
Note that although our server has two CPUs, we use one CPU and its corresponding memory in the following experiments.
Thus, the experimental results do not include the effect of non-uniform memory access.

\begin{table}[b]
  \caption{Experimental environment.}
  \label{tab:environment}
  \centering
  \begin{tabular}{ll}
    \toprule
    Item     & Value                                            \\
    \cmidrule(r){1-1}
    \cmidrule(l){2-2}
    CPU      & Intel Xeon Gold 6258R (two sockets)              \\
    RAM      & DIMM DDR4 Registered 2933 MHz (16GB $\times$ 12) \\
    PMEM     & Intel Optane 100 series (128GB $\times$ 8)       \\
    OS       & Ubuntu 20.04.2 LTS                               \\
    Compiler & GNU C++ version 9.4.0                            \\
    PMDK     & Version 1.8                                      \\
    \bottomrule
  \end{tabular}
\end{table}

\begin{figure}[b]
  \centering
  \includegraphics{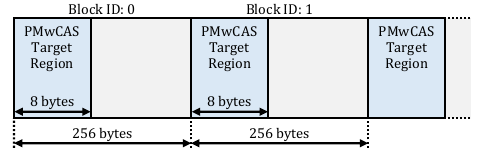}
  \caption{Memory blocks for benchmarking PMwCAS operations.}
  \label{fig:memory-blocks}
\end{figure}

We use two competitors to evaluate the proposed methods.
\begin{enumerate}
  \item \textbf{Wang et al.'s PMwCAS} (denoted as ``Original'' in figures):
        This is a reference implementation~\cite{URL:microsoft-PMwCAS} used in the original paper~\cite{ICDE:Wang2018}.
        Note that since the implementation has a bug related to flush instructions, we fixed it before benchmarking.
  \item \textbf{A software PCAS} (denoted as ``PCAS'' in figures):
        We implemented Wang et al.'s PCAS operation~\cite{ICDE:Wang2018} in C++.
        In this implementation, we use TTAS and back-off~\cite{BOOK:Herlihy2021ch7} to avoid frequent cache invalidation and flushes, as described in \Sec{\ref{sec:pmwcas-with-df}}.
        We use the software PCAS as a competitor for persistent one-word CAS operations to show ideal throughput and latency.
\end{enumerate}
In the following figures, we use ``Ours (DF)'' and ``Ours'' to denote the proposed methods with/without dirty flags.

The benchmarking program has the following features.
\begin{itemize}
  \item It prepares one million words $W$ as candidates for PMwCAS operations.
        We allocate a memory block for each word and use the head 8 bytes as a target region of PMwCAS operations, as shown in \figref{fig:memory-blocks}.
        Each memory block has 256 bytes as defaults to simulate the components of data structures (e.g., leaf/inner nodes in \BTree{s}).
        It also avoids false sharing~\cite{BOOK:Herlihy2021appB} in cache lines and Intel Optane persistent memory.
  \item Each PMwCAS operation selects the specified number of words randomly.
        We follow the existing benchmarks~\cite{SOCC:Cooper2010,TOS:Yang2016} and use a parameter $\alpha$ to control the skew of the $k$-th word to be selected according to Zipf's law:
        \begin{equation}
          f(k; \alpha, |W|) = \frac{1 / k^{\alpha}}{\sum_{n = 1}^{|W|} 1 / n^{\alpha}}.
        \end{equation}
        In the experiments, we used $\alpha = 0$ and $\alpha = 1$ as low/high-competitive environments, respectively.
  \item Every PMwCAS operation reads the current value of each target word and adds one atomically.
        If a PMwCAS operation fails, it \emph{continues} its own processing until it succeeds.
  \item Each worker thread executes at most one million PMwCAS operations with a timeout.
        When worker threads reach a specified timeout, they interrupt their processing and compute throughput/latency based on completed PMwCAS operations.
        We used ten seconds as a timeout.
\end{itemize}

We ran the benchmarking program five times and measured the average throughput or latency.
The following graphs contain no error bars because the measurement results are sufficiently stable.
Note that we used all the threads in a CPU (56 threads) to measure each latency and performed sampling to compute percentile latency.
Thus we use one and ninety-ninth percentile latency instead of the minimum and maximum latency, respectively.

\begin{figure*}[t]
  \centering
  \includegraphics{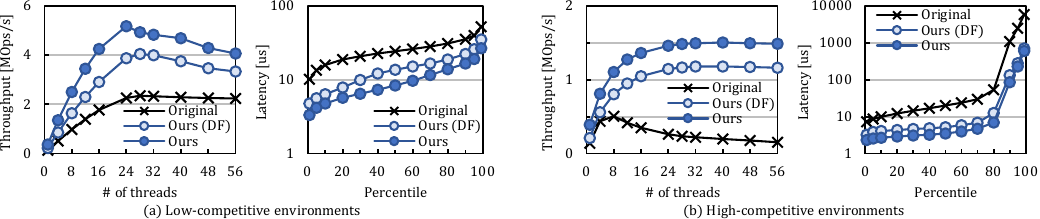}
  \caption{Comparison on throughput and percentile latency of persistent three-word CAS operations.}
  \label{fig:exp-3wCAS}

  \vspace*{6truemm}

  \centering
  \includegraphics{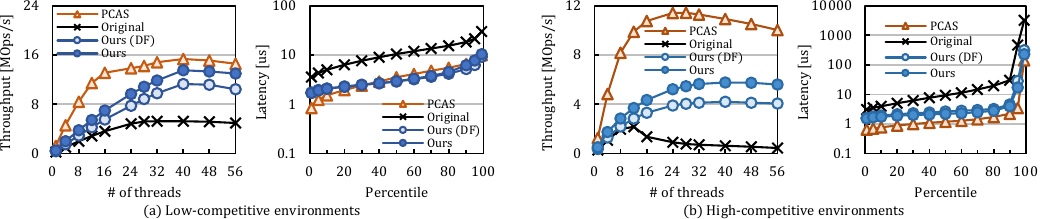}
  \caption{Comparison on throughput and percentile latency of persistent one-word CAS operations.}
  \label{fig:exp-1wCAS}

  \vspace*{6truemm}

  \centering
  \includegraphics{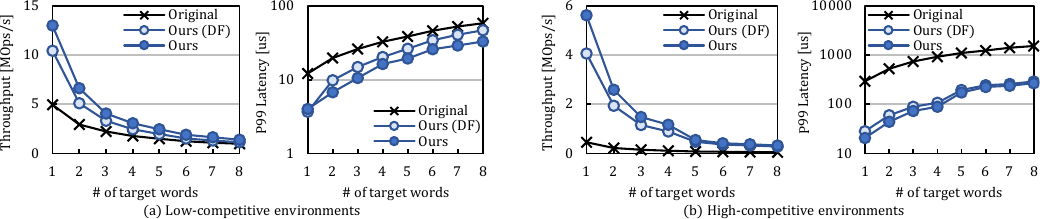}
  \caption{Comparison on throughput and 99th percentile latency with different numbers of target words.}
  \label{fig:exp-over-target}
\end{figure*}

\subsection{Effects of Excluding Redundant Cache Invalidations}

First, we demonstrate the performance improvements of the proposed methods due to avoiding redundant cache invalidations and flushes.
In this experiment, we use persistent three/one-word CAS operations in low/high-competitive environments with different numbers of worker threads.

\figref{fig:exp-3wCAS} shows the throughput and latency of the competitors with persistent three-word CAS operations.
In low-competitive environments, our methods achieve twice the throughput and half the latency of the original PMwCAS algorithm.
Since the conflicts of PMwCAS rarely happen in low-competitive environments, this result shows the fundamental efficiency of the proposed methods.
In high-competitive environments, the proposed methods achieve ten times the throughput and a tenth the latency of the original PMwCAS.
As Wang et~al.'s algorithm performs redundant CAS and flush instructions when happening the conflict of PMwCAS operations, the throughput of their method decreases with many cores.
On the other hand, our methods avoid such useless CAS/flush instructions and keep the throughput over different numbers of threads.
The differences between our methods with/without dirty flags also show the inefficiency of frequent flush instructions and the effectiveness of removing dirty flags.

\figref{fig:exp-1wCAS} shows the throughput and latency of the competitors with persistent one-word CAS operations.
In low-competitive environments, our methods achieve comparable throughput and latency with the software PCAS.
Since the PCAS and the proposed method without dirty flags require the same number of CAS instructions (i.e., setting/clearing a dirty flag or embedding/removing a descriptor) for a single PMwCAS, their fundamental efficiency is similar.
In high-competitive environments, on the other hand, our methods have half the throughput and twice the latency of the software PCAS.
These results show the limitation of using our approach for persistent one-word CAS operations.
While the proposed method without dirty flags needs double flushes of PMwCAS target addresses to guarantee consistency, the PCAS algorithm can guarantee consistency with a single flush.
This difference in the number of flushes affects the performance of one-word CAS operations, and these results also show the importance of avoiding frequent flushes.

\subsection{Effects of Parameters}

We evaluate the performance trends of the proposed methods using three parameters: the number of target words, the skew parameter of a Zipf distribution, and the memory block size of each word.
We use the 99th percentile latency to evaluate latency in the following experiments.

\subsubsection{Effects of the Number of Target Words}

\figref{fig:exp-over-target} shows the throughput and 99th percentile latency with different numbers of PMwCAS target words in low/high-competitive environments.
These results show that the proposed methods outperform Wang et~al.'s PMwCAS algorithm with any number of target words.
In particular, our approach achieves at least six times the throughput of the existing method in high-competitive environments because of the reduction of CAS/flush instructions.

\begin{figure}[tb]
  \centering
  \includegraphics{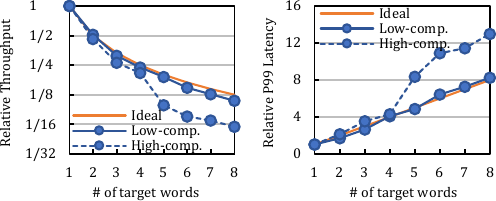}
  \caption{The relative throughput and 99th percentile latency in relation to P1wCAS operations with different numbers of target words.}
  \label{fig:exp-over-target-relative}
\end{figure}

On the other hand, the performance of all methods decreases as the number of target words increases.
Since the number of target words is directly related to the number of executions of CAS and flush instructions, this performance trend is inevitable.
For example, \figref{fig:exp-over-target-relative} shows the relative throughput and 99th percentile latency in relation to persistent one-word CAS operations of our PMwCAS without dirty flags, and the ``Ideal'' line shows the desired relative throughput/latency.
That is, since persistent two-word CAS operations have twice as many targets as persistent one-word CAS operations, its throughput will be half the throughput of persistent one-word CAS even in the ideal case.
As shown in \figref{fig:exp-over-target-relative}, the performance degradation as a function of the number of target words is reasonable in low-competitive environments.
However, in high-competitive environments, PMwCAS performance degrades with five or more target words due to its long critical sections.

\subsubsection{Effects of Access Skew}

\begin{figure*}[tb]
  \centering
  \includegraphics{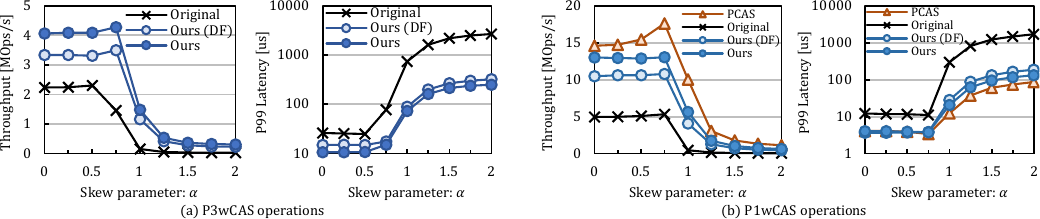}
  \caption{Comparison on throughput and 99th percentile latency with different values of the skew parameter.}
  \label{fig:exp-over-skew}

  \vspace*{6truemm}

  \centering
  \includegraphics{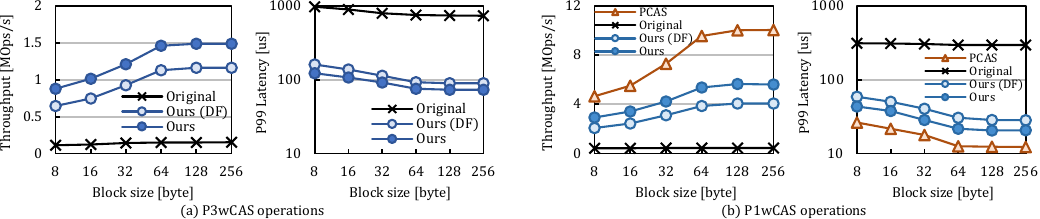}
  \caption{Comparison on throughput and 99th percentile latency with different values of the block size.}
  \label{fig:exp-over-block}
\end{figure*}

\figref{fig:exp-over-skew} shows the throughput and 99th percentile latency of persistent one/three-word CAS operations with different values of the skew parameter.
These results show that the proposed methods outperform the original PMwCAS algorithm regardless of access skew.
Although the proposed method without dirty flags have half the throughput and twice the latency of the software PCAS in high-competitive environments, the results in low-competitive environments show that the fundamental efficiency of the proposed method is comparable to that of the PCAS operations.

As shown in the results, the effect of access skew is similar across different numbers of target words.
In low-competitive environments (i.e., $\alpha < 1$), since the conflicts of PMwCAS operations occur infrequently, all the methods achieve stable throughput and latency.
As environments become high-competitive (i.e., $\alpha \geq 1$), worker threads tend to access the same target addresses for their PMwCAS operations and cause conflicts.
Since worker threads are required to resolve these conflicts (e.g., wait for other threads to finish their PMwCAS operations) and read flushed data from persistent memory, PMwCAS performance is degraded.

\subsubsection{Effects of False Sharing}

\figref{fig:exp-over-block} shows the throughput and 99th percentile latency of persistent one/three-word CAS operations with different values of the memory block size.
For example, if we use 64 bytes as the memory block size, at most four words can be in the same read/write unit of Intel Optane (i.e., 256 bytes).
On the other hand, since each word does not share its memory region in CPU cache lines (64 bytes), false sharing in CPU caches does not occur.
Note that we only use high-competitive environments in this experiment to evaluate the effects of false sharing.

These results show that false sharing in persistent memory does not affect the performance of PMwCAS operations.
Since Intel Optane uses 256 bytes as its read/write unit, 64/128-byte block sizes cause false sharing in persistent memory.
However, its effect is negligible because Intel Optane has a write buffer to avoid redundant writes into the same write units.

On the other hand, false sharing in CPU cache lines significantly degrades throughput and latency.
As the current CPUs use 64 bytes as the size of cache lines, 8/16/32-byte block sizes cause false sharing in cache lines.
False sharing in cache lines causes frequent cache invalidations and redundant flush instructions of shared lines.
Thus, as shown in \figref{fig:exp-over-block}, the 8-byte block size has half the throughput and twice the latency of 64-byte block size.

\subsection{Discussion}

This subsection summarizes three suggestions for using PMwCAS operations based on the above experimental results.

\textbf{1. Keep the number of PMwCAS target words as small as possible.}
Although PMwCAS operations help developers with atomic and durable word swapping, swapping too many words will cause non-negligible performance degradation.
It tends to occur in high-competitive environments, as shown in \figref{fig:exp-over-target-relative}, and PMwCAS operations typically have such conflict-sensitive words as targets.
Thus, developers should reduce target words as much as possible to achieve high-performance multithreaded software.

\textbf{2. Avoid placing multiple words on the same CPU cache lines.}
As shown in \figref{fig:exp-over-block}, false sharing in CPU cache lines significantly degrades PMwCAS performance.
Therefore, developers should design their data structures to contain at most one PMwCAS target word in each cache line.
For example, since \BTree{s} typically have their control words (e.g., locks) in each node header, a node header should be 64 bytes long and be aligned to cache lines.

\textbf{3. If there is a high-competitive PMwCAS target word, swap it first.}
As shown in the experimental results above, cache invalidation due to CAS and flush instructions is one of the critical problems in improving PMwCAS performance.
If developers place a high-competitive word last and it fails, \emph{PMwCAS operations must revert their process while invalidating the corresponding cache lines}.
Although swap order is also related to the linearization of PMwCAS operations, developers should design procedures to swap high-competitive words as early as possible.

\section{Conclusion}
\label{sec:conclusion}

In this paper, we proposed the new PMwCAS algorithms for many-core CPUs.
We reduced CAS and flush instructions in PMwCAS operations and excluded dirty flags using PMwCAS descriptors as write-ahead logs.
The experimental results demonstrated the efficiency of the proposed methods; our PMwCAS without dirty flags is up to ten times faster than the original algorithm.
Future work includes improving a persistent \BTree{}, BzTree~\cite{PVLDB:Arulraj2018}, using our PMwCAS implementation.

\begin{acknowledgment}
  This paper is based on results obtained from a project, JPNP16007, commissioned by the New Energy and Industrial Technology Development Organization (NEDO).
  In addition, this work was partly supported by JSPS KAKENHI Grant Numbers JP20K19804, JP21H03555, and JP22H03594.
\end{acknowledgment}

\bibliographystyle{ipsjsort-e}
\bibliography{reference}

\end{document}